\documentclass[aps,prl,twocolumn,superscriptaddress,amsmath,amssymb]{revtex4-1}

\usepackage{graphicx}
\usepackage{multirow}
\usepackage{natbib}
\usepackage{array}
\usepackage{multirow,amssymb,amsbsy,amsmath,epstopdf}
\usepackage{color}

\usepackage[colorlinks,citecolor=blue]{hyperref}

\begin{document}
\title{Demonstration of Multi-Setting One-Way Einstein-Podolsky-Rosen Steering in Two-Qubit Systems}

\author{Ya Xiao}
\affiliation{CAS Key Laboratory of Quantum Information, University of Science and Technology of China, Hefei 230026, People's Republic of China}
\affiliation{Synergetic Innovation Center of Quantum Information and Quantum Physics, University of Science and Technology of China, Hefei 230026, People's Republic of China}

\author{Xiang-Jun Ye}
\affiliation{CAS Key Laboratory of Quantum Information, University of Science and Technology of China, Hefei 230026, People's Republic of China}
\affiliation{Synergetic Innovation Center of Quantum Information and Quantum Physics, University of Science and Technology of China, Hefei 230026, People's Republic of China}

\author{Kai Sun}
\affiliation{CAS Key Laboratory of Quantum Information, University of Science and Technology of China, Hefei 230026, People's Republic of China}
\affiliation{Synergetic Innovation Center of Quantum Information and Quantum Physics, University of Science and Technology of China, Hefei 230026, People's Republic of China}

\author{Jin-Shi Xu}\email{jsxu@ustc.edu.cn}
\affiliation{CAS Key Laboratory of Quantum Information, University of Science and Technology of China, Hefei 230026, People's Republic of China}
\affiliation{Synergetic Innovation Center of Quantum Information and Quantum Physics, University of Science and Technology of China, Hefei 230026, People's Republic of China}

\author{Chuan-Feng Li}\email{cfli@ustc.edu.cn}
\affiliation{CAS Key Laboratory of Quantum Information, University of Science and Technology of China, Hefei 230026, People's Republic of China}
\affiliation{Synergetic Innovation Center of Quantum Information and Quantum Physics, University of Science and Technology of China, Hefei 230026, People's Republic of China}

\author{Guang-Can Guo}
\affiliation{CAS Key Laboratory of Quantum Information, University of Science and Technology of China, Hefei 230026, People's Republic of China}
\affiliation{Synergetic Innovation Center of Quantum Information and Quantum Physics, University of Science and Technology of China, Hefei 230026, People's Republic of China}
\date{\today}

\begin{abstract}
Einstein-Podolsky-Rosen (EPR) steering describes the ability of one party to remotely affect another's state through local measurements. One of the most distinguishable properties of EPR steering is its asymmetric aspect. Steering can work in one direction but fail in the opposite direction. This type of one-way steering, which is different from the symmetry concepts of entanglement and Bell nonlocality, has garnered much interest. However, an experimental demonstration of genuine EPR steering in the simplest scenario, i.e., one that employs two-qubit systems, is still lacking. In this work, we experimentally demonstrate one-way EPR steering with multimeasurement settings for a class of two-qubit states, which are still one-way steerable even with infinite settings. The steerability is quantified by the steering radius, which represents a necessary and sufficient steering criterion. The demonstrated one-way steering in the simplest bipartite quantum system is of fundamental interest and may provide potential applications in one-way quantum information tasks.

\end{abstract}

\maketitle



In 1935, Einstein, Podolsky and Rosen (EPR) described a ``spooky'' action permissible under the rules of quantum mechanics: ``as a consequence of two different measurements performed upon the first system, the second system may be left in states with two different (ensembles of) wave functions''  \cite{EPR1935}. As a response to EPR's work, Schr\"{o}dinger generalized this argument   and referred to the ability of Alice to remotely affect Bob's state by choosing her measurement basis as steering \cite{Schrodinger1935,Schrodinger1936}. The rigorous definition and operational framework for understanding steering were recently formulated by Wiseman \emph{et al.} \cite{Wiseman2007,Jones2007}, in which the authors showed the hierarchy of nonlocality: steerable states are a subset of the entangled states and a superset of Bell nonlocal states \cite{Bell1964}.

Another interesting property of steering according to the definition is its asymmetry: Alice and Bob play different roles in the steering scenario. For a given two-party system, one can ask whether Alice can steer Bob, which shows Alice's ability to remotely affect Bob's states and vice versa. This formal asymmetry can never be found in entanglement or Bell nonlocality by their definitions, which may provide potential applications for the one-sided device-independent quantum key distribution \cite{Branciard2012,Opanchuk2014,Walk2016}.

It is natural to verify steering by violating steering inequalities. However, to certify a one-way steerable state, one needs to solve two obstacles. The first difficulty is when all those one-way steerable states are Bell-local states \cite{Wiseman2007}; thus, a highly efficient and experimental error-tolerant steering criterion is required to verify Alice's ability to steer Bob. The second difficulty, which is the most challenging part, is to prove, for any measurement settings, that Bob cannot steer Alice. Great efforts have been made in designing one-way EPR steering tests. The asymmetry of the EPR-steering correlation was first investigated by Wiseman \emph{et al}. They offered the problem of whether there exists an asymmetric quantum steering state as the foremost open question in their work \cite{Wiseman2007}. Later, it was shown theoretically \cite{Olsen2008,Midgley2010,Olsen2013} that such a phenomenon could occur in continuous variable systems. However, these results hold only for a restricted class of measurements, i.e., Gaussian measurements, and there was no evidence that this asymmetry would persist for more general measurements. A year later, Bowles \emph{et al.} theoretically confirmed for the first time that quantum nonlocality can be fundamentally asymmetric \cite{Bowles2014}. They presented a class of one-way steerable states in a two-qubit system with at least 13 projective measurements. However, the requirement for the state preparation is very high, and it is difficult to experimentally demonstrate the corresponding steerability. Recently, they further investigated the one-way steering problem by presenting a sufficient criterion (a nonlinear criterion) for guaranteeing that a two-qubit state is unsteerable \cite{Bowles2016}, which provides a general method for constructing the one-way steerable states.

A few experiments have also been carried out over the past few years to study the asymmetric steering. The first experimental demonstration was restricted to Gaussian measurements for Gaussian states \cite{Handchen2012}. Recently, two more experiments were conducted to observe the one-way steering. Based on the analysis of detector efficiency, Wollmann \emph{et al.} \cite{Wollmann2016} designed an experiment to demonstrate one-way steering in a qubit-qutrit system, which consisted of a Werner state with a lossy channel at one side \cite{Evans2014,Skrzypczyk2014}. However, because the dimensionality of the prepared state is asymmetric, one may doubt where the one-way steering characteristic comes from. The lossy channel, which introduces an additional vacuum state on one side, is essential to their protocol. Without this increased dimension, the asymmetric steering cannot be demonstrated in principle.
The other experiment regarding asymmetric EPR-steering was reported by Sun \emph{et al.} \cite{Sun2016}, in which the protocol is restricted to two-measurement settings. The one-way steering characteristic may disappear with more measurement settings. The demonstration of genuine one-way steering with multimeasurement settings in the simplest bipartite system would be of fundamental interest and provide practical applications in one-way quantum information tasks, which are still lacking. This suggests that further experimental efforts are needed.

In this work, we experimentally demonstrate the multi-setting one-way steering in a two-qubit system for the first time. The steerability is quantified by a necessary and sufficient steering criterion, i.e., the steering radius \cite{Sun2016}. These results will provide a deeper understanding of the asymmetric characteristic of steering.



\begin{figure}[!htb]
\begin{center}
\includegraphics[width=0.85\columnwidth]{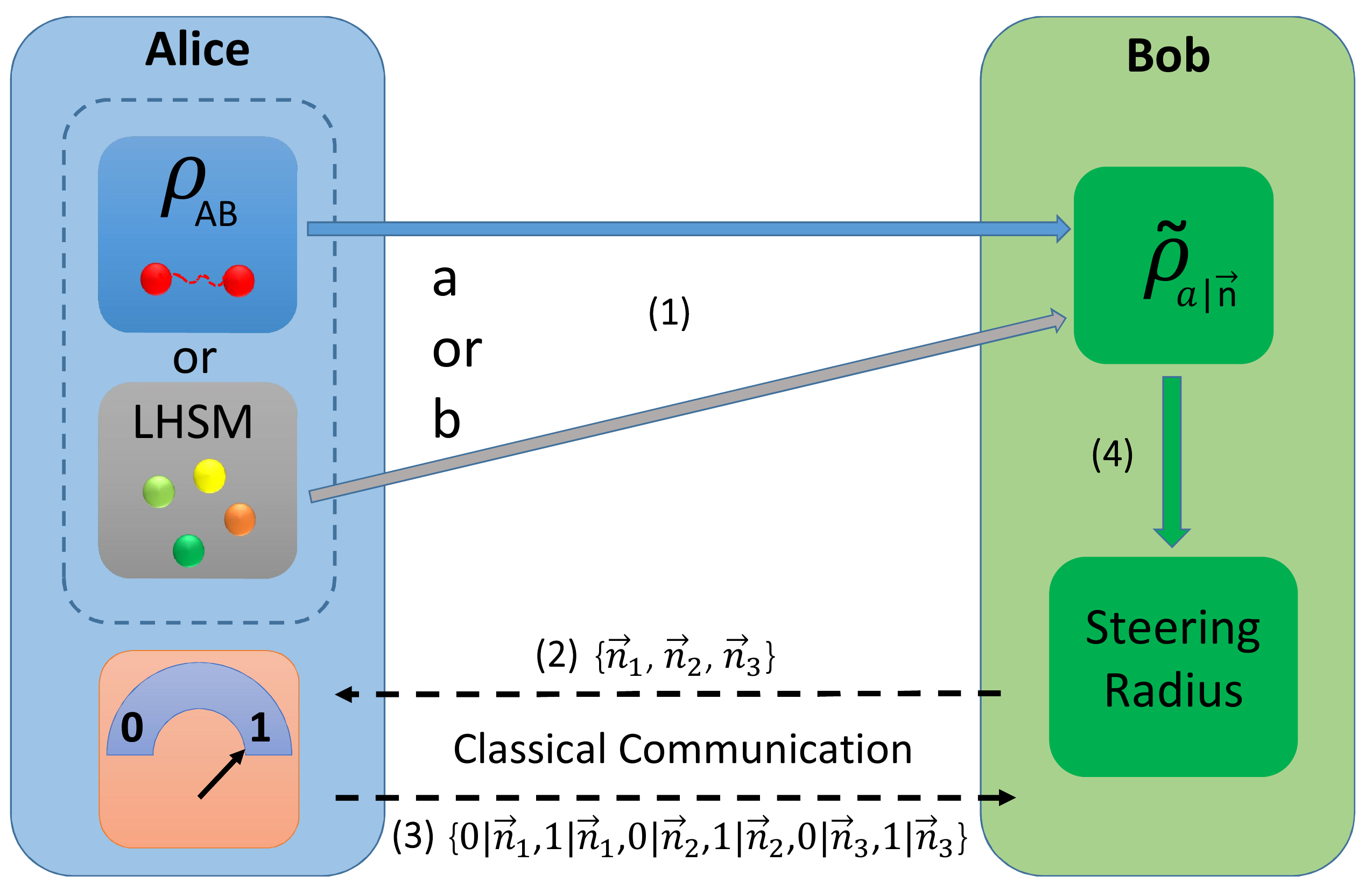}
\caption{Illustration of the EPR steering scenario with three-measurement settings. The steps in the task, from (1) to (4), are as follows. (1) Alice sends a qubit to Bob, who is not sure whether the state is from a steerable state ($\rho_{AB}$, channel a) or from a local hidden state model (LHSM, channel b). (2) Alice measures her qubit along one of the directions $\lbrace \vec{n}_{1},\vec{n}_{2},\vec{n}_{3}\rbrace $ required by Bob through classical communication. (3) Alice sends her measurement results $\lbrace 0\vert\vec{n}_{1}, 1\vert\vec{n}_{1}, 0\vert\vec{n}_{2},1\vert\vec{n}_{2},0\vert\vec{n}_{3},1\vert\vec{n}_{3} \rbrace $ to Bob. Six corresponding conditional states are obtained by Bob, which are denoted as $ \tilde{\rho}_{a\vert\vec{n}} $, with $ a\in \lbrace 0,1\rbrace $ and $ \vec{n}\in \lbrace\vec{n}_{1},\vec{n}_{2},\vec{n}_{3}\rbrace $. (4) Bob analyzes the results to calculate the steering radius $\it{R_{A\rightarrow B}}$. If $ R_{A\rightarrow B}>1 $, Alice successfully steers Bob's state. The qubit received by Bob is confirmed to be from channel a.}

\label{stmodel}
\end{center}
\end{figure}

\textit{Multi-setting one-way EPR steering and steering radius}.---To clarify the steering scenario, we show the process of Alice steering Bob in the case of three-measurement settings in Fig. \ref{stmodel}. Bob is not sure whether the received qubit state is from half of a steerable state ($\rho_{AB}$, from channel a) or from a local hidden state model (LHSM, from channel b). He asks Alice to measure her qubit in one of the directions $\lbrace \vec{n}_{1},\vec{n}_{2},\vec{n}_{3}\rbrace $ through classical communication. Alice then sends Bob the measurement result {\it a} $\in \{0, 1\}$. Correspondingly, Bob obtains a conditional state $ \tilde{\rho}_{a\vert\vec{n}}=Tr_{A}( (M_{a\vert\vec{n}} \otimes I_{B})\rho_{AB})$.  $ M_{a\vert\vec{n}}=(I_{A}+(-1)^a\vec{n}\cdot\vec{\sigma})/2 $ is the measurement operator of Alice's state. {\it $I_{A}$} ({\it $I_{B}$}) represents the identity matrix on Alice's (Bob's) side, and $ \vec{\sigma}= (\sigma_{x},\sigma_{y},\sigma_{z})$ is the Pauli vector. When Bob's state can be described by a LHSM, the conditional state can be written as

\begin{equation}\label{local}
\tilde{\rho}_{a\vert\vec{n}}=\sum \limits_{i} P{(a\vert\vec{n},i)}p_{i}\rho_{i},
\end{equation}
where $ P{(a\vert\vec{n},i)}$ is a conditional probability. $\lbrace p_i\rho_i\rbrace$ is the local hidden state ensemble, with the state $\rho_i$ and corresponding probability $p_i$. Otherwise, Bob is convinced that Alice can steer his state and that the qubit he received is from channel a.


Here, we use a value called steering radius $\it{R_{A\rightarrow B}}$ to quantify the ability of Alice to steer Bob \cite{Sun2016}. In the case of three-measurement settings, it has been proven that eight local hidden states are sufficient to reproduce the six conditional states if a LHSM exists \cite{Wu2014}. The radius of the Bloch vector of the corresponding local hidden state $ \rho_i $ is represented as $ \vert \vec{R}_{i}\vert $, with $ i\in\lbrace 1,2 \cdots 8\rbrace $. For a different solution set of $\lbrace p_i\rho_i\rbrace$ of Eq. \ref{local}, the minimum radius is defined as $ r_{\lbrace \vec{n}_{1},\vec{n}_{2}, \vec{n}_{3}\rbrace}=\min \limits_{\lbrace p_{i}\rho_{i}\rbrace}{\lbrace max \lbrace \vert \vec{R}_{i}\vert \rbrace \rbrace} $. Obviously, $ r_{\lbrace \vec{n}_{1},\vec{n}_{2}, \vec{n}_{3}\rbrace } $ is dependent on a given measurement direction assemblage $ {\lbrace \vec{n}_{1},\vec{n}_{2},\vec{n}_{3}\rbrace } $. The steering radius is defined as $ R_{A\rightarrow B}=\max \limits_{\lbrace \vec{n}_{1},\vec{n}_{2},\vec{n}_{3}\rbrace }\lbrace{r_{\lbrace \vec{n}_{1},\vec{n}_{2},\vec{n}_{3}\rbrace}}\rbrace $. If $ R_{A\rightarrow B}>1 $, there is no physical solution of Eq. \ref{local}, which indicates that there is no LHSM to describe the conditional states obtained on Bob's side. The steering task from Alice to Bob is successful. Otherwise, if $ R_{A\rightarrow B}\leq 1 $, the EPR steering task fails. The analysis can be extended to more measurement settings, and the steering radius for the case in which Bob steers Alice ($R_{B\rightarrow A}$) can be analyzed in a similar way.


In this work, we prepare a family of two-qubit states:
\begin{equation}\label{states}
 \rho_{AB}(p,\theta)=p \vert \psi(\theta)\rangle\langle \psi(\theta)\vert+(1-p)I_{A}/2\otimes\rho_{B}^{\theta},
\end{equation}
where $  \vert\psi(\theta)\rangle =\cos(\theta)\vert HH\rangle+\sin(\theta)\vert VV\rangle$, with {\it H} and {\it V} representing the horizontal and vertical polarizations, respectively. $ \rho_{B}^{\theta} = Tr_{A} (\vert \psi(\theta)\rangle\langle \psi(\theta)\vert) $. It has been demonstrated \cite{Bowles2016}  that for $ \theta \in [0,\pi/4] $ and $ \cos^{2}(2\theta)\geqslant\dfrac{2p-1}{(2-p)p^{3}} $, the steering from Bob to Alice is impossible even for an infinite number of projective measurements carried out by Alice. However, Alice can steer Bob for $ p>1/2 $.


Experimentally, we focus on two- and three-measurement settings. The conditions of states $ \rho_{AB} $ satisfying one-way steering from Alice to Bob are $ R_{A\rightarrow B}>1$ and $ R_{B\rightarrow A}\leq 1 $. In the case of two-measurement settings, the condition can be rewritten as:
$ \theta \in (0,\pi/4) $ and
\begin{equation}\label{cirt2}
\dfrac{1}{\sqrt{2}}< p\leq \dfrac{1}{\sqrt{1+\sin^{2}(2\theta)}}.
\end{equation}
while the condition for three-measurement settings is
$ \theta \in (0,\pi/4) $ and
\begin{equation}\label{cirt3}
\dfrac{1}{\sqrt{3}}< p\leq \dfrac{1}{\sqrt{1+2\sin^{2}(2\theta)}}.
\end{equation}
The detailed calculation and proof are shown in the Supplementary Material  \cite{SM}.





\begin{figure}[!htb]
\begin{center}
\includegraphics[width=1\columnwidth]{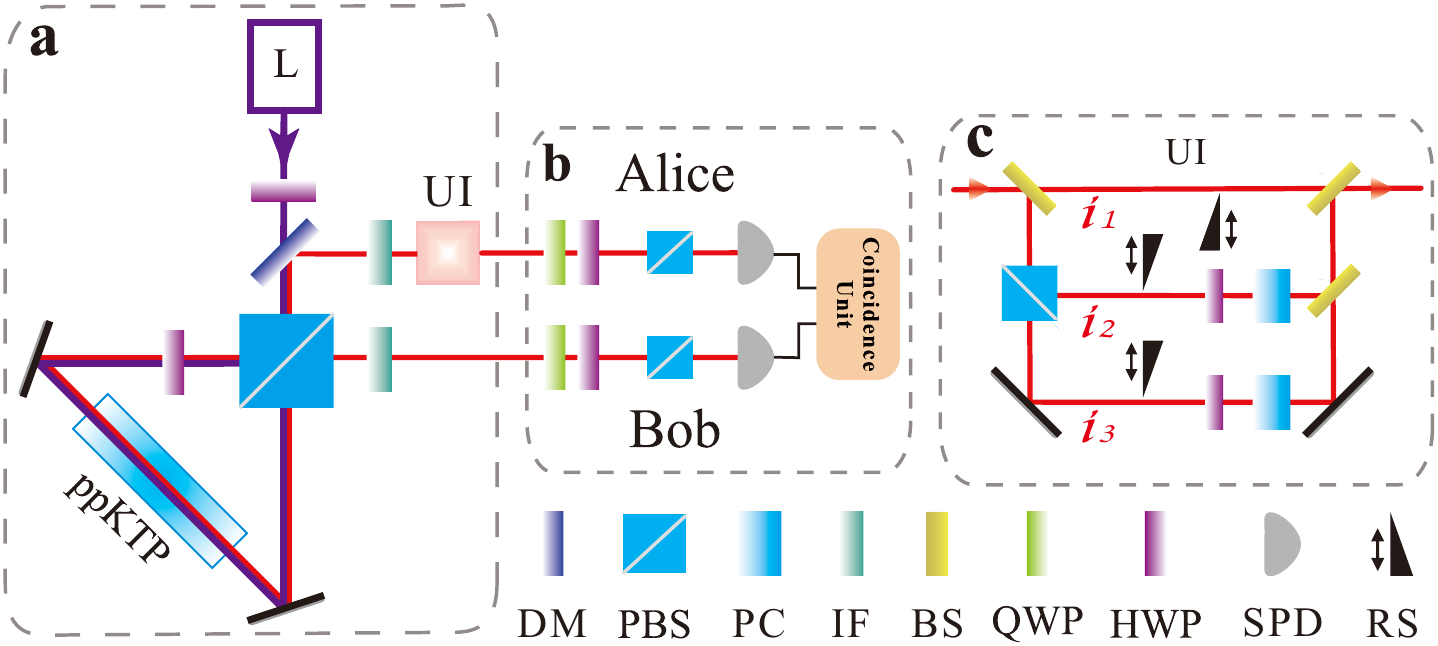}
\caption{Experimental setup. \textbf{a. State Preparation.} A pair of photons in a state $ \vert\psi(\theta)\rangle$ is generated via the spontaneous parametric down-conversion process by pumping a type-II cut PPKTP crystal located in a Sagnac interferometer with an ultraviolet laser (L) at 404 nm. The parameter $\theta$ is controlled by the half-wave plate (HWP) in front of the laser. These two photons are filtered by interference filters (IFs). One of the photons passes through an unbalanced interferometer (UI) for state preparation and is sent to Alice. The other photon is sent to Bob. {\bf b. Measurement settings.} The polarization analyzer consisting of a quarter-wave plate (QWP), a HWP and a polarization beam splitter (PBS) on both sides of Alice and Bob is used for measurement settings. The photons are detected by single-photon detectors (SPDs), and the signals are sent for coincidence. {\bf c. The unbalance interferometer (UI).} The state of path $  i_{1} $ remains unchanged. HWPs set at $ 22.5^{\circ} $ and two sufficiently long birefringent crystals (PCs) along paths $  i_{2} $ and $  i_{3} $ introduce a sufficiently large time delay between $ \vert H\rangle $ and $ \vert V\rangle  $ components, which can completely destroy the coherence. By combining these three paths into one, arbitrary two-qubit states $ \rho_{AB}(p,\theta)$ can be prepared. The parameter $ p $ can be controlled conveniently by employing removable shutters (RSs).}

\label{setup}
\end{center}
\end{figure}

\emph{Experimental setup and results}.---Fig. \ref{setup} shows our experimental setup. A 404 nm continuous-wave diode laser (L) with polarization set by a half-wave plate is used to pump a 20 mm-long PPKTP crystal inside a polarization Sagnac interferometer \cite{Fedrizzi2007} to generate polarization-entangled photons in the state $ \vert\psi(\theta)\rangle$. Two interference filters with a bandwidth of 3 nm are used to filter the photons. One of the two photons is sent to an unbalanced interferometer (UI) and then sent to Alice. The other photon is sent to Bob directly. In the UI, the photon is separated into three paths, denoted as $i_{1}, i_{2} $ and $  i_{3} $, by a beam splitter (BS) and a polarization beam splitter (PBS). The state of path $i_{1}$ remains unchanged. Half-wave plates (HWPs) along paths $ i_{2} $ and $  i_{3} $ are set at $ 22.5^{\circ} $, and two sufficiently long birefringent crystals (PCs) introduce a sufficiently large time delay between $ \vert H\rangle $ and $ \vert V\rangle $ components, which can completely destroy the coherence. The time difference between these three paths is much larger than the coherence time of the photons. By combining these three paths into one, arbitrary two-qubit states $ \rho_{AB}(p,\theta) $ can be prepared. The parameter $ p $ can be controlled conveniently by employing removable shutters (RSs).

The measurement setup, comprising a quarter-wave plate (QWP), a HWP, and a PBS on both sides 
allows us to measure along arbitrary axes on the Bloch sphere for each qubit. For two- and three-measurement settings, according to the symmetrical property of the steering ellipsoid of $\rho_{AB}(p,\theta)$ \cite{Jevtic2014,Jevtic2015}, the optimal choice of measurement settings is $\lbrace \vec{x}, \vec{z}\rbrace $ and $\lbrace \vec{x}, \vec{y}, \vec{z}\rbrace $ for both steering directions, respectively. When the measurement is carried out on one side, the other will obtain the corresponding conditional states. Then, we can check whether the state is one-way steerable by evaluating the steering radii $\it{R_{A\rightarrow B}}$ and ${\it R_{B\rightarrow A}}$.


We prepared 40 entangled states in the form of $ \rho_{AB}(p,\theta) $ to perform the EPR steering task. The detailed process for determining the experimental parameters $ p $ and $ \theta $ is shown in the Supplementary Material  \cite{SM}.
Fig. \ref{steerP3}\textbf{a} presents the distribution of the experimental states with different $ p $ and $ \theta $. In the scenario of three-measurement settings, the light red region described by Ineq. \ref{cirt3} denotes the case of one-way steering in which Alice can steer Bob, but Bob cannot steer Alice. In other cases, states located in the light brown region are steerable, and states located in the light blue region are unsteerable in both directions.
It is clear that a tunable $ p $ allows the state to be shifted from a region where it is unsteerable in both directions to a region where it is one-way steerable and finally to a region where it is two-way steerable. We further show the one-way steering region in the case of two-measurement settings, which is bounded by the dashed black lines according to Ineq. \ref{cirt2}. With more measurement settings, more states are shown to be steerable. One-way steerable states can be turned into two-way steerable states by increasing the measurement settings for some parameters. For the infinite-measurement settings, there is still a parameter region where states are shown to be one-way steerable, which could not be demonstrated in the previous work restricted to two-measurement settings \cite{Sun2016}. As shown in Fig. \ref{steerP3}\textbf{a}, the states below the solid red curve described by the relation $ \cos^{2}(2\theta)=\dfrac{2p-1}{(2-p)p^{3}} $ are one-way steerable with infinite-measurement settings  \cite{Bowles2016}.

We further consider the ability of Bob to steer Alice using the linear EPR-steering inequality, which is represented as $ S_{n}=\dfrac{1}{n} \sum^{n}_{k=1} \langle \sigma_{A}^{k} B^{k}\rangle \leqslant C_n $. $\sigma_A^k$ is the Pauli operator for Alice's state, and $B^k\in \{-1, 1\}$ is the random variable on Bob's side. $C_n$ is the bound given by the LHSM with {\it n}-measurement settings. The difference between $S_n$ and $C_{n}$ ($S_{n}-C_{n}$) for the one-way steerable states in the black box in Fig. \ref{steerP3}\textbf{a} is shown in Fig. \ref{steerP3}\textbf{b}. If $S_{n}-C_{n}>0$, the steerability is demonstrated. We find that $S_{n}-C_{n}$ slowly increases as the number of measurement settings increases. All $S_n$ are shown to be well below $C_n$.

However, $S_{n}-C_{n}$ is not a necessary and sufficient criterion to quantify steering for states $\rho_{AB}(p,\theta)$. It only shows, for a specific linear function, whether there exists a LHSM to obtain the value predicted by quantum mechanics. Thus, it just tests partial properties of the conditional states. However, the steering radius, as a simplified variant of steering robustness~\cite{Piani2015}, which is discussed in Ref.~\cite{Sun2016}, directly shows whether there exists a LHSM to simulate the corresponding conditional states and gives a necessary and sufficient criterion for steerability. We measure the steering radii $R_{A\rightarrow B}$ and $R_{B\rightarrow A}$ of the corresponding states in Fig. \ref{steerP3}{\bf a} to clearly demonstrate the one-way EPR steering, as shown in Fig. \ref{steerR3}.
The blue dots represent states for which the EPR steering task fails in both directions ({\it A}$ \nleftrightarrow ${\it B}, $R_{A\rightarrow B}\leq1$ and $R_{B\rightarrow A}\leq1$). The states represented by red triangles show the case in which Alice steers Bob ({\it A}$\rightarrow ${\it B}, $R_{A\rightarrow B}>1$ and $R_{B\rightarrow A}\leq1$). The brown squares represent the cases in which Alice and Bob can steer each other ({\it A}$\leftrightarrow ${\it B}, $R_{A\rightarrow B}>1$ and $R_{B\rightarrow A}>1$). The values of $ R_{A\rightarrow B} $ and $ R_{B\rightarrow A} $ clearly distinguish different steering situations, which agree well with the theoretical predictions. The inset in Fig. \ref{steerR3} is a magnification of the region in the red pane. Error bars are due to Poissonian counting statistics. The experimental states and steering radius in the case of two-measurement settings are discussed further in the Supplementary Material \cite{SM}.


\begin{figure}[!htb]
\begin{center}
\includegraphics[width=\columnwidth]{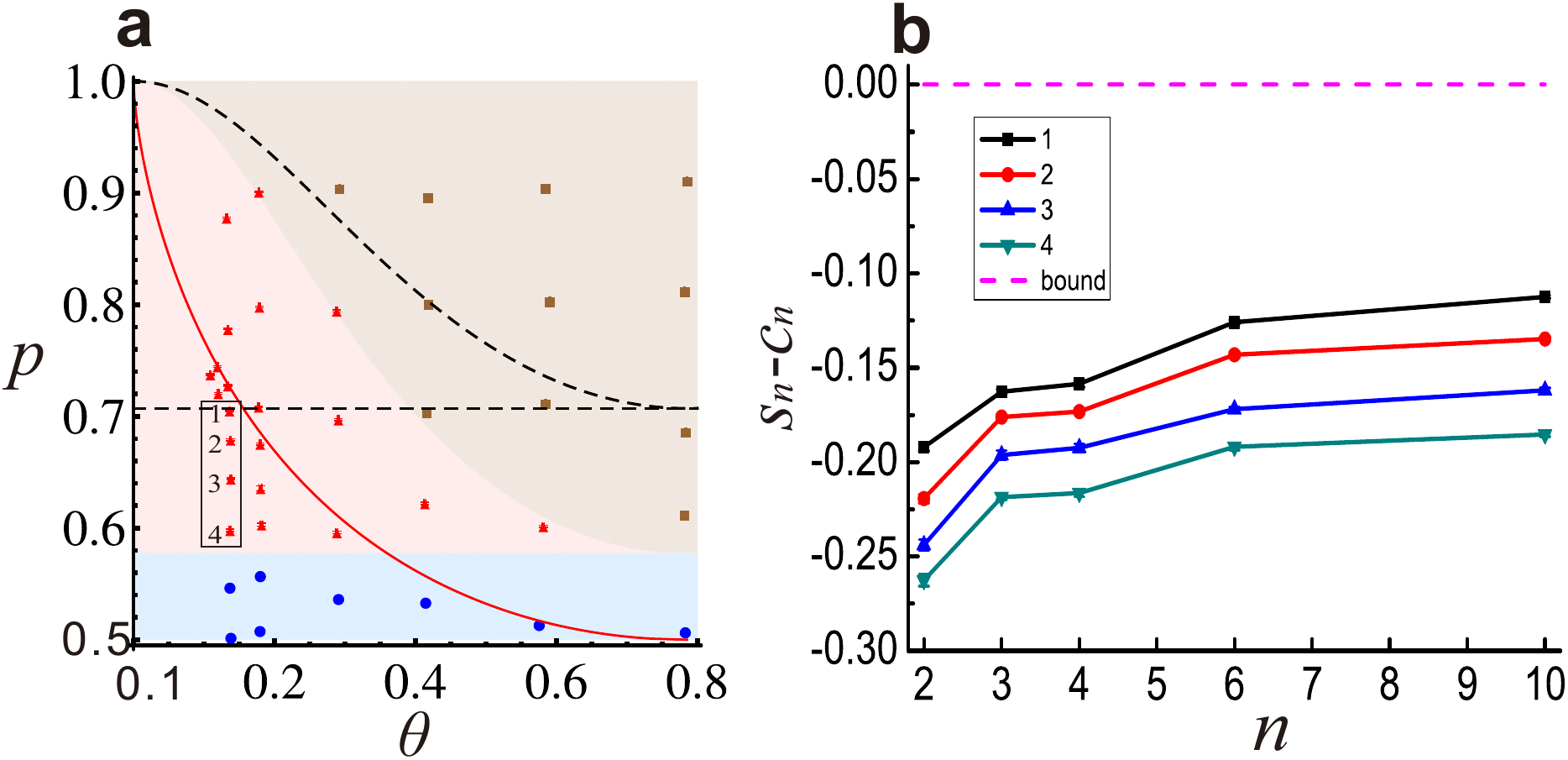}
\caption{Distribution of experimental states and the value of $ S_n-C_n $. \textbf{a}. The distribution of experimental states. In the case of three-measurement settings, the light blue region represents the states for which the steering task fails in both directions. The light red region represents the case of one-way steering, i.e., Alice can steer Bob's state, but Bob cannot steer Alice's state. The light brown region represents the case for which Alice and Bob can steer each other. The blue dots, red triangles and brown squares in the corresponding regions are the experimental states. The area between two dashed black curves corresponds to the one-way steerable states in the case of two-measurement settings. States below the solid red curve are one-way steerable for infinite measurement settings. \textbf{b}. The results of $ S_n-C_n $ of the four one-way steerable states are shown in the black pane in {\bf a}. All $ S_n $ are smaller than $ C_n $. Error bars are due to the Poissonian counting statistics, which are small, i.e., within the sizes of the symbols.}

\label{steerP3}
\end{center}
\end{figure}


\begin{figure}[!htb]
\begin{center}
\includegraphics[width=0.8\columnwidth]{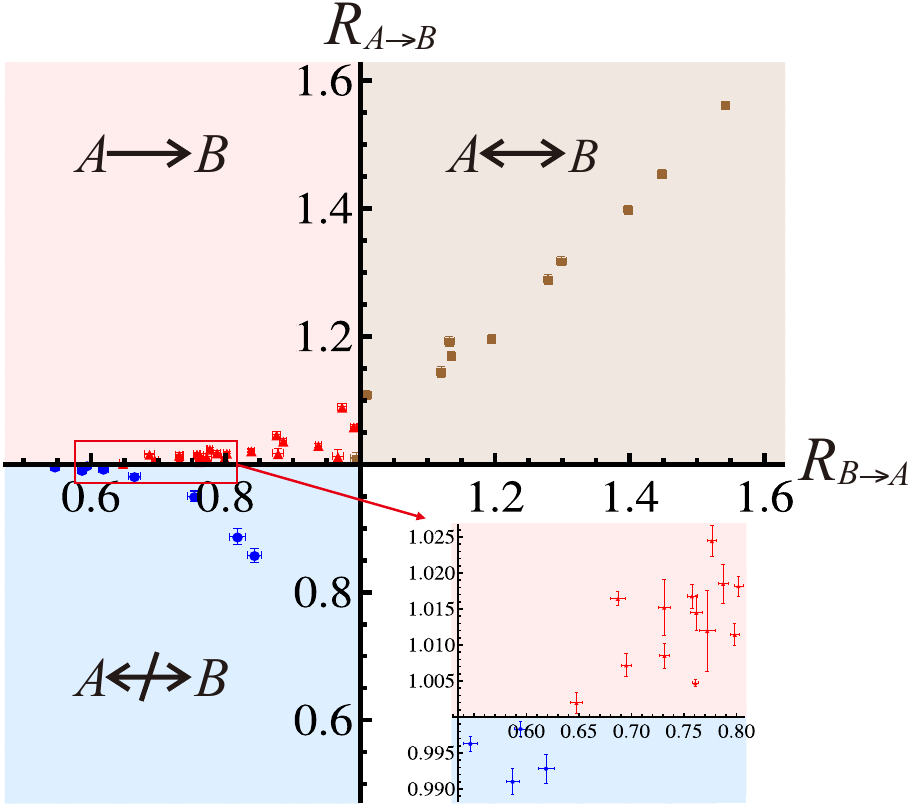}\label{steerR3}
\caption{The values of $R_{A\rightarrow B}$ and $R_{B\rightarrow A}$. Different steering situations are clearly distinguished by the values of $  R_{A\rightarrow B} $ and $ R_{B\rightarrow A} $, i.e., two-way steerable ({\it A}$ \leftrightarrow ${\it B}, brown squares), one-way steerable ({\it A}$ \rightarrow ${\it B}, red triangles) and unsteerable ({\it A}$ \nleftrightarrow ${\it B}, blue dots). The inset is the magnification of the corresponding region in the red pane. Error bars are due to the Poissonian counting statistics.}

\label{steerR3}
\end{center}
\end{figure}



\textit{Conclusion.}---In our work, we construct a class of states that are only steerable from Alice to Bob, even for infinite-measurement settings. By measuring the steering radius, the asymmetric steerability of the prepared states is clearly shown. Compared with the previous experiments, our work provides a more essential and intuitive way to understand the asymmetric characteristic of EPR steering. Our experimental results for the simplest bipartite system, with a smaller requirement of quantum resources, can yield potential applications in future one-way quantum information tasks.

This work was supported by National Key Research and
Development Program of China (Grants No. 2016Y-FA0302700), the National Natural Science Foundation of China (Grant Nos. 61327901, 11325419, 61322506, and 11274297), the Strategic Priority Research Program (B) of the Chinese Academy of Sciences (Grant No. XDB01030300), the Key Research Program of Frontier Sciences, CAS (Grants No. QYZDY-SSW-SLH003), the Fundamental Research Funds for the Central Universities (Grant No. WK2470000020) and the Youth Innovation Promotion Association and Excellent Young Scientist Program CAS.

Y. X. and X.-J. Y. contributed equally to this work.

\end{document}